# Role of Digital Twin in Optical Communication: Fault Management, Hardware Configuration, and Transmission Simulation

Danshi Wang, Zhiguo Zhang, Min Zhang, Meixia Fu, Jin Li, Shanyong Cai, Chunyu Zhang, and Xue Chen

*Abstract*—Optical communication is developing rapidly in the directions of hardware resource diversification, transmission system flexibility, and network function virtualization. Its proliferation poses a significant challenge to traditional optical communication management and control systems. Digital twin (DT), a technology that utilizes data, models, and algorithms and integrates multiple disciplines, acts as a bridge between the real and virtual worlds for comprehensive connectivity. In the digital space, virtual models are established dynamically to simulate and describe the states, behaviors, and rules of physical objects in the physical space. DT has been significantly developed and widely applied in industrial and military fields. This study introduces the DT technology to optical communication through interdisciplinary crossing and proposes a DT framework suitable for optical communication. The intelligent fault management model, flexible hardware configuration model, and dynamic transmission simulation model are established in the digital space with the help of deep learning algorithms to ensure the high-reliability operation and high-efficiency management of optical communication systems and networks.

## INTRODUCTION

Currently, the world is entering the digital era, where digital twins (DTs) have become an anticipated enabling technology to promote digital transformation and intelligent evolution in various areas [1]. DT, which creates high-fidelity digital virtual models of physical objects to simulate their behaviors and depict their operating states, paves the way for realizing cyber–physical fusion [2]. The power of advanced computing and analytics in the cyber space opens a bright perspective to all walks of life. DT is receiving increasing attention and has been used in a wide range of fields, including the manufacturing industry, military engineering, marine engineering, aerospace engineering, electrical power systems, and smart cities [3, 4].

Owing to its high bandwidth, low loss, and strong anti-interference, optical fiber communication is the key technology to realize large-capacity and long-haul data transmission in this era of big data. Ensuring a stable operation, as well as efficient and intelligent management, of optical communication systems and networks is significant and indispensable. However, with the proliferation of the elastic optical transceivers, complex optical modulation, hybrid optical transmission, variable wavelength grids, software-defined optical networks, and other technologies, optical communication is rapidly evolving in the direction of hardware resource diversification, transmission system flexibility, and network function virtualization [5, 6]. Moreover, owing to the complex deployment environments and wide range of coverage, the existing optical communication systems and networks encounter severe challenges in their accurate monitoring, on-line management, real-time control, and rapid maintenance [7]. In addition, the conventional network management schemes always take no account of actual physical situation. In researches of optical network, there is a premise that the conditions of physical layer are ideal, stable, and static, and the characteristics of physical layer are mainly based on the simplified numerical models, but always ignore the real physical world. Therefore, there is a yawning gap between network layer and physical layer. Although DT has been used widely in various fields, it has not yet been applied to optical communication. In view of the technical challenges faced in optical communication, DT has the potential to provide low-cost and lightweight solutions to address these challenges, and it may have broad application prospects in optical communication.

Considering the need to develop optical communication, and encouraged by the technical advantages of DT, this study introduces

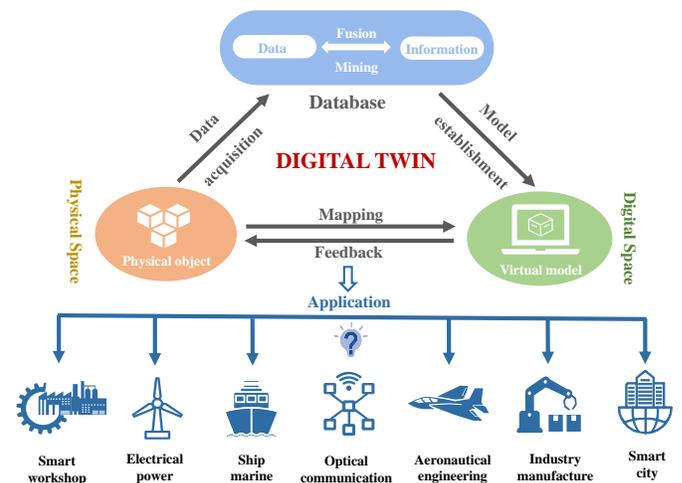

Fig. 1. The basic principle and application background of digital twin.

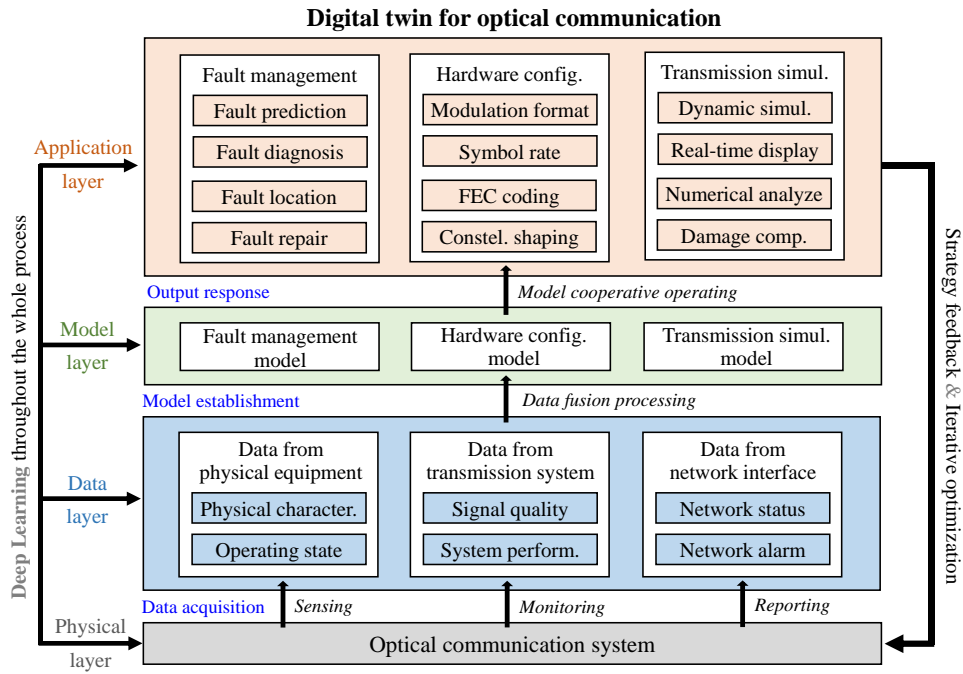

Fig. 2. Digital twin framework for optical communication consisting of physical layer, data layer, model layer, and application layer.

DT to optical communication to perform multiple functions, including fault management, hardware configuration, and transmission simulation. First, a DT framework is designed particularly for optical communication. In the physical space, the acquisition and fusion of multi-source heterogeneous data from physical objects are implemented using a cognitive optical network. In the digital space, an intelligent fault management model, a flexible hardware configuration model, and a dynamic transmission simulation model are established. Deep learning as a powerful interdisciplinary tool plays a central role in the entire DT system to realize all types of functions, including data acquisition, model building, result analysis, and strategy execution. The application of the DT technology in optical communication is realized to ensure a safe and stable operation of optical communication systems, optimize the hardware efficiency and network resource efficiency, and improve the dynamic monitoring and numerical analysis capability of the transmission system.

DIGITAL TWIN FRAMEWORK FOR OPTICAL COMMUNICATION

The basic principle and application background of DT are depicted in Fig. 1. In the physical space, real-time data acquisition from physical objects is achieved through sensing and monitoring. The collected data are processed and stored in a database. After data mining and fusion processing, useful information is obtained and deep-level knowledge is explored. Based on the collected data and captured information, virtual models can be built dynamically in the digital space. According to various application requirements, various virtual models generate the corresponding target results and feedback the matching optimization strategies to the physical space. The reality–virtual interaction is realized by forward mapping to virtual models and backward reaction to physical objects. Based on the idea of DT, a series of successful applications has been demonstrated in the fields of smart workshops, electric power, ship navigation, aeronautical engineering, manufacturing industry, and smart cities (see Fig. 1). It has implemented the various services, such as intelligent manufacturing, work flow control, operation status prediction, lifetime monitoring, system simulation, numerical analysis, 3D visualization, etc.

Next, we carried out exploratory research on combination of DT and optical communication. Firstly, the framework of DT particularly designed for optical communication is proposed, which is composed of physical layer, data layer, model layer, and application layer, as shown in Fig. 2. From the bottom up, the physical layer in this architecture refers to the physical objects in optical communication system, i.e., optical equipment, network elements, transmission modules, fiber links, etc. Based on the idea of cognitive optical network, through sensor detection, optical performance monitoring, and network message reporting, all sorts of data can be collected in real time from physical equipment, transmission systems, and network interface sources. Then the acquired data are sent and stored in the data layer, where the fusion processing and deep mining of multi-source heterogeneous data are carried out. Both the unified data after fusion processing and the useful information obtained from data mining are subsequently transmitted to the model layer.

Here, we mainly focus on the three essential subjects in optical communication: fault management, hardware configuration, and transmission simulation. Therefore, in the model layer, we investigate and build the intelligent equipment fault management model, flexible hardware configuration model, and dynamic transmission process simulation model simultaneously. Through the cooperative operation and dynamic response, these three models can perform the corresponding functions in the application layer, including the equipment fault prediction/diagnosis/location/repairment, hardware resource optimization and network efficiency improvement, transmission process dynamic simulation, real-time display, and numerical analysis. Finally, according to the comprehensive analysis results, the optimization strategies are formulated and response actions are taken in the application layer, and then feedback to the physical layer from the top down. Through reality-virtual interaction between forward mapping and backward reaction, the DT system for optical communication forms a closed loop. As a supporting technology, deep learning algorithm runs through the whole process to implement data acquisition, model establishment, numerical analysis, and strategy execution [8]. Eventually, a framework of DT specialized for optical communication system is established.

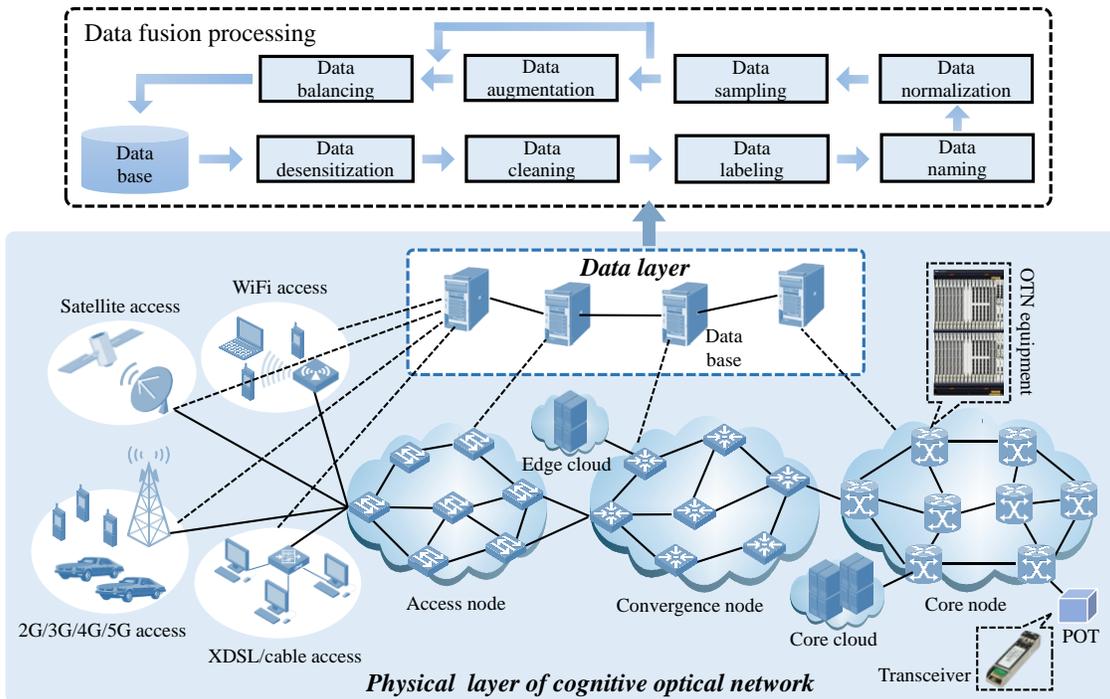

Fig. 3. Multi-source heterogeneous data collected from cognitive optical network consisting of physical layer of cognitive optical network and data layer for data fusion processing.

## DATA ACQUISITION

The realization of DT is based on the massive data pertaining to all elements and over the entire lifetime. A virtual model can be accurately built in the digital space only when sufficient data are collected from the physical space. In optical communication, right from the physical layer to the network layer, significant amounts of data are generated from a wide range of sources, including historical data, initial data, and real-time update data for the network status, equipment operating state, and transmission system performance. All types of multi-source heterogenous data are acquired from cognitive optical network [9], which aims to introduce cognition on multiple planes (e.g., access, aggregation, and core nodes) to perceive current network conditions, as shown in Fig. 3. The data on the network status (e.g., resource utilization information, notification message, delay jittering, and blocking rate) are obtained from the network log viewer. The data on the equipment operating state (e.g., environment temperature, unusable time, input/output optical power, and laser bias current) are acquired from sensing detectors. The data on the performance of the transmission system (e.g., optical signal-to-noise ratio (OSNR), modulation format, transmission speed, and system impairment) are obtained from optical performance monitoring (OPM) modules. In addition, the transmitted signal can also be obtained by photoelectric detector. However, different data have different structural characteristics, operation modes, storage mechanisms, and matching algorithms. Therefore, it is essential to implement data fusion for all the multi-source heterogeneous data in the data layer. Data fusion refers to data desensitization, cleaning, labeling, naming, normalization, sampling, augmentation, and balancing, which convert the original data into processable data and usable information, as shown in Fig. 3.

Based on the data type, multi-source heterogeneous data can be broadly divided into three categories: image data, time-series data, and other structural data. All these categories must be comprehensively fused and further processed using deep learning. Various deep learning algorithms are available for this purpose. In terms of the algorithm mechanism and data category, the convolutional neural network, recurrent neural network, deep neural network, and their variants can be used to process image data, time-series data, and structured data [10, 11]. In addition, depending on research objects and modeling requirements, deep learning can be used to extract important features and capture deep-level information, such as the prediction of equipment operating parameters, estimation of the optical signal quality, evaluation of the performance of the transmission system, and analysis of the network status. All these are of significance for the establishment of a model.

## MODEL ESTABLISHMENT

DT establishes a quasi-real-time connection between the physical and digital worlds through virtual models. It uses modeling technology to describe the physical state, simulate the operation process, forecast change tendency, and optimize object performance. Therefore, it can be said that the model layer is the most central part of the DT system. However, the traditional modeling methods in optical communication suffer from the following disadvantages: (a) strong dependence on expert experience and mathematical theories, (b) effective only for the static and single scenario, and (c) inability to perform real-time iterative optimization. While in digital twin, massive real-time data are collected and stored in the data layer. A wealth of information and a variety of time-varying characteristics are contained in the various data. Different from traditional model-driven methods, in our scheme, based on the idea of real-time data-driven modeling, three deep learning-based models are proposed: one each for intelligent fault management, flexible hardware configuration, and dynamic transmission simulation.

## FAULT MANAGEMENT MODEL

Large-scale optical networks consist of plenty of optical equipment. A failure of the network can cause serious consequences, such as data loss, transmission interruption, and performance deterioration. The traditional approach to the fault management of the optical network equipment is based on the built-in inspection software provided by the manufacturer to monitor some typical operating parameters. When the real-time monitoring values reach the alarm threshold value, the fault management system automatically sends out an alarm. This type of fault management is simple and straightforward, and its technical architecture is relatively mature. However, parameter monitoring and threshold settings are generally based on experience and lack precision. Furthermore, the traditional fault management system is capable of only a broad fault alarm and fault prediction and lacks an effective fault diagnosis function. Therefore, there is a need for not only improving the accuracy but also extending the function scope of fault management.

With the help of DT, optical network equipment equipped with sensing and monitoring modules can collect various operating state parameters in real time, such as the environment temperature, laser power, bias current, etc. Based on this historical operating state data obtained from the equipment in the physical space, virtual fault prediction and fault diagnosis models are generated in the digital space using two learning algorithms. First, a bidirectional gated recurrent unit (BiGRU) algorithm is used to predict the trend of the equipment state parameters that can be regarded as temporal sequences. Next, the predicted state parameters are used to prognosticate whether the equipment will fail at a certain time.

At present, we have demonstrated the feasibility of this scheme on optical transport network (OTN) equipment (ZXOME 8300) those are mainly deployed in the metropolitan area network [12]. The data from preceding thirty-six days are collected for model training. Based on the trained model, the operating state of the following seven days can be predicted. It can predict the occurrence of the fault at least one day in advance, with an average accuracy of greater than 99% and a false rate of less than 0.9%.

Next, extreme gradient boosting (XGBoost) is selected to implement fault diagnosis [13]. XGBoost is an integrated learning algorithm that adopts the classification and regression tree (CART) as the base learner. Unlike traditional neural networks, the base learner of XGBoost is composed of a root node, branches, and leaf nodes. The path from the root node to the leaf node can explain the information of a certain decision rule. Encouraged by the idea that XGBoost finds the best split through feature splitting continuously, it is possible to introduce XGBoost to interpret the cause of the fault and reveal the important features for fault diagnosis. The greater the number of times the feature node splits, the stronger the correlation between the feature node and fault. Based on the data from operating state records of existing OTN, there are total 15 features that may cause equipment fault. By decomposing the internal structure of XGBoost and analyzing node characteristics, the 5 most remarkable features responsible for the fault are found, as shown in Fig. 4. The feature scoring dependent on split time can be calculated in XGBoost. In terms of their importance grades, the possible fault causes are the laser bias current, environment temperature, output optical power, input optical power, and laser temperature offset in order, which can provide valuable reference information for fault repair.

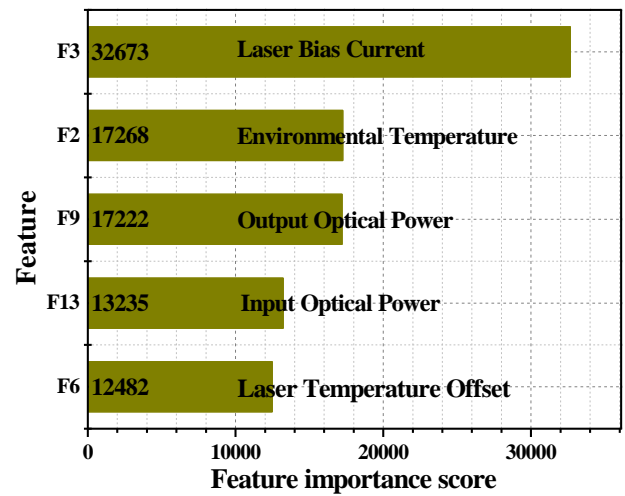

Fig. 4. Top 5 feature importance in XGBoost feature scoring for fault diagnosis.

## HARDWARE CONFIGURATION MODEL

In dynamic optical networks, programmable optical equipment serves as one of the key equipment that can flexibly adjust the configuration parameters to meet the diverse quality of service levels and adapt to the time-varying network environment. In this study, we use the programmable optical transceiver (POT) to investigate the application of DT in the hardware configuration. As the core network equipment, POT can flexibly configure multi-dimensional parameters, such as the modulation format (MF), symbol rate (SR), forward error correction (FEC) coding, probabilistic constellation shaping, and subcarrier, depending on the real-time link state, optical signal quality, and service transmission requirements for the purpose of flexible bandwidth resource allocation and automatic transmission capacity regulation. The ideal POT configuration model can learn the nonlinear mapping relationship between the POT operating parameters and the corresponding transmission performances under various network conditions to provide the optimal configuration strategy quickly. However, the classical POT configuration model is mainly dependent on expert experience, lacks a feedback mechanism, and is incapable of a dynamic response. Therefore, it fails to meet the requirements of dynamic optical networks. Encouraged by the strong dynamic digital modeling and intelligent feedback controlling characteristics of DT, we propose a DT-enabled POT adaptive configuration scheme using deep reinforcement learning (DRL), which is oriented by low delay and high spectral efficiency.

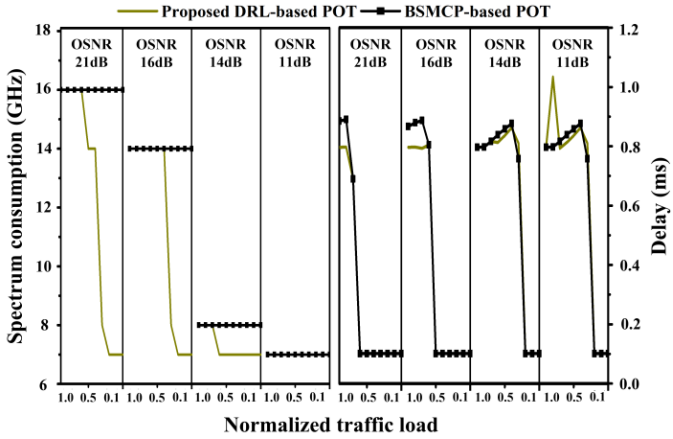

Fig. 5. Comparisons between the proposed DRL-based scheme and the BSMCP-based scheme in terms of the spectrum consumption and the network delay.

Based on the operating principle of DT, we propose a self-evolving POT based on DRL [14]. In the physical space, POT provides transmission bandwidth resources flexibly to satisfy the diverse quality of service levels between the source and destination nodes. Through configuring multi-dimension parameters of POT, the spectrum consumption and delay performance are affected accordingly. In the data base, multidimensional monitoring data on POT collected from the physical space are uploaded to the data fusion platform for data desensitization, cleaning, and normalization. The data fusion platform comprises multi-dimensional monitoring data and information, including but not limited to OSNR and control actions from physical layer, transmission capability and network delay from network layer, and bandwidth and delay requirements from application layer. After data fusion, the POT monitoring data are input to the digital space.

After fusion processing, the POT monitoring data are sent to the digital space, where DRL is used to implement the dynamic digital modeling and devise the smart control strategy for the POT configuration. In the DRL agent, the double deep Q-learning network consisting of two deep neural networks—modeling and evolving networks with similar structures—is selected. The objective of the modeling network is to reduce network delay and improve the utilization of the spectrum. Through the historical and real-time POT control experience, the modeling network is set to determine the uncertain relationship between the network state and transmission performance utilization under various POT control actions. In contrast, the evolving network is used to learn real-time POT control experience and, accordingly, adjust the parameters of the modeling network to adapt to the dynamic network environment. This scheme can achieve a balance between transmission quality and spectrum resource cost; that is, it can significantly improve the utilization of the spectrum and reduce the network delay while meeting the transmission requirements for various services in various physical link transmission states.

The experiment and simulation results indicate that, when compared with the classical POT based on brute-force search and maximum capability provisioning (BSMCP), the POT based on DT-DRL can dynamically adjust the parameters of the control model depending on the environmental changes, ensure the service transmission requirements, and improve the utilization of the spectrum. In the Fig. 5, the spectrum occupation decreases with the normalized traffic load and less spectrum consumption can be obtained in the proposed DRL-based scheme, where the average spectrum consumption is about 77.9% of those of the BSMCP-based scheme. Meanwhile, the average network delay performance curves are almost overlapped in two schemes where the average delay in the proposed scheme is 0.7% larger than those in BSMCP-based scheme. We are of the view that DT will open a new avenue for adaptive optical component modeling and control of dynamic optical networks.

## TRANSMISSION SIMULATION MODEL

Performing simulations of optical communication systems is essential for system designs. It helps to simulate the actual operating process and characterize the real environment numerically. However, the traditional simulation systems for optical communication are usually composed of a series of model blocks, and each model block is established based on rigorous mathematical theory. Therefore, traditional simulation systems are generally effective only for static and ideal transmission scenarios and are not valid for practical transmission scenarios (where the link characteristics and signal performance vary dynamically in real time) and the newly evolving transmission scenarios whose analytical models have not yet been accurately proposed and mathematically verified. Using DT and deep learning, we propose a data-driven dynamic modeling technique for simulating the process of optical transmission instead of block-based system, as shown in Fig. 6. According to the characteristics of deep learning, the model functions can be approximated by mapping independent to dependent variables corresponding to the input and output data. Deep learning could construct an approximate channel model driven by source data and received data. In the physical space, signal data before transmission and after reception are collected in real time. Deep learning-based dynamic response models are established in the digital space and driven by the collected data without dependence on experience and mathematical theory.

In the physical space, the optical transmission system consists of an optical transmitter, optical channel, optical amplifier, and optical receiver. In the digital space, the transmission process simulation presents the entire process of optical signal transmission from the transmitter, through fiber links, after relay amplifications to the receiver. Therefore, the optical signal data from transmission and reception contain the entire real-time information, such as the signal quality, channel characteristics, and equipment performance, of the dynamic transmission process. In [15], we have demonstrated a data-driven fiber channel modeling using deep learning to simulate the transmission process through a standard single-mode fiber. First, optical signals from the optical transmitting module and receiving modules are collected as the input and output data to train the deep learning model. The input and output optical signals are composed of amplitude and phase information, which are typical time-series data. Considering the correlation between the front and rear sequences, the multilayer bidirectional long short-term memory (BiLSTM) is selected as the deep learning algorithm because of its superior performance for temporal sequences. Based on the amplitude and phase information of the optical signals, two BiLSTM networks are used to execute amplitude and phase waveform fitting. Based on the real-time collected amplitude and phase data of optical signals, BiLSTM can effectively learn the transfer function and realize waveform fitting through iterative optimization training to depict the complete transmission process of the optical signal from its transmission to its reception in the digital space. Through testing and verification, we have realized a 0–80 km optical channel simulation model for the intensity modulation and direct detection system [15], which proves the feasibility of the data-driven modeling scheme. Furthermore, because the scheme does not rely on expert experience, it can significantly reduce the modeling cost and improve the simulation efficiency. This transmission simulation model in DT system can not only digitize the physical process, but also provide the numerical channel model that is important for adaptive damage compensation, like end-to-end learning method, to ensure high reliable transmission of optical communication.

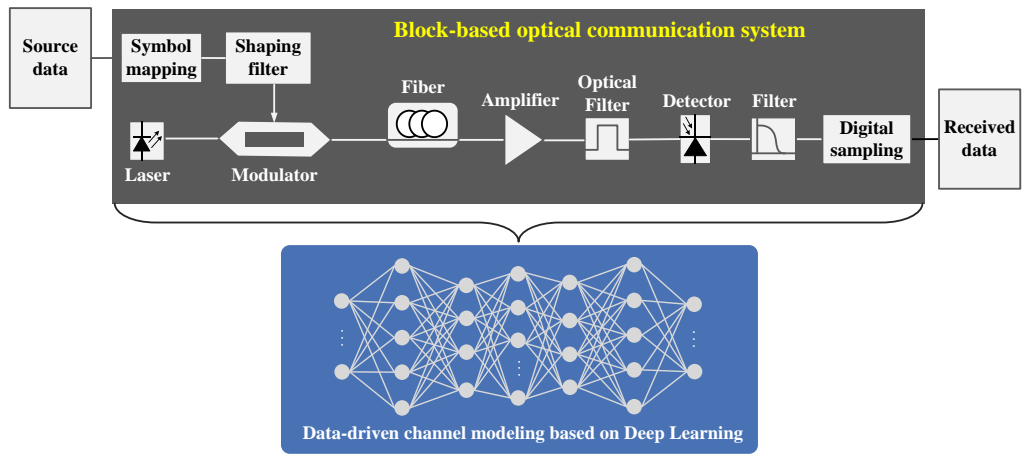

Fig. 6. Data-driven channel modeling based on deep learning to perform transmission simulation instead of block-based optical communication system constructed in a divide-and-conquer manner.

## CONCLUSIONS

In this study, we investigated the role of DT technology in optical communication and discussed its application for fault management, hardware configuration, and transmission simulation. A DT framework suitable for optical communication was designed, the problem of mass data acquisition and fusion in the physical space was analyzed, and multiple deep learning-based virtual models were proposed. DT technology can effectively improve and ensure the control efficiency and stable operation of optical communication systems and promote their digital transformation and intelligent evolution. In addition to the three types of application scenarios discussed in this study, DT can also be used to realize other potential applications, such as the lifecycle management and 3D visualization of optical equipment, adaptive damage compensation of transmission systems, intelligent monitoring of network state and system performance, and flexible allocation of network resources, in the future.

## ACKNOWLEDGEMENT

This work was supported by National Natural Science Foundation of China (No.61871415, No.61671076, No. 61705016) and Fund of State Key Laboratory of IPOC (BUPT) (No. IPOC2020ZT05), P. R. China.

## REFERENCES

[1] F. Tao, J. Cheng, Q. Qi, M. Zhang, H. Zhang, and F. Sui, "Digital twin-driven product design, manufacturing and service with big data," *The Inter. J. Adv. Manufac. Technol.*, vol. 21, no. 9, pp. 3563-3576, 2018.
[2] E. Negri, L. Fumagalli, and M. Macchi, "A Review of the Roles of Digital Twin in CPS-Based Production Systems," *Procedia Manufacturing*, vol. 11, pp. 939-948, 2017.
[3] E. J. Tuegel, A. R. Ingraffea, T. G. Eason, and S. M. Spottswood, "Reengineering aircraft structural life prediction using a digital twin," *Intern. J. Aero. Eng..*, pp. 1687-5966, 2011.
[4] F. Tao, H. Zhang, A. Liu, and A. Y. Nee, "Digital Twin in Industry: State-of-the-Art," *IEEE Trans. Indus. Inform.*, vol. 15, no.4, pp. 2405-2415, 2019.
[5] Z. Zhu, W. Lu, L. Zhang, and N. Ansari, "Dynamic Service Provisioning in Elastic Optical Networks with Hybrid Single-/Multi-Path Routing," *IEEE J. Lightw. Technol.*, vol. 31, no. 1, pp. 15-22, Jan. 2013.
[6] P. Lu, L. Zhang, X. Liu, J. Yao, and Z. Zhu, "Highly-Efficient Data Migration and Backup for Big Data Applications in Elastic Optical Inter-Data-Center Networks," *IEEE Netw.*, vol. 29, no. 5, pp. 36-42, 2015.
[7] F. Musumeci *et al.*, "An overview on application of machine learning techniques in optical networks," *IEEE Commun. Surv. Tut.*, vol. 21, no. 2, pp. 13831408, 2018.
[8] F. N. Khan, Q. Fan, C. Lu, and A. P. T. Lau, "An Optical Communication's Perspective on Machine Learning and Its Applications," *J. Lightwave Technol.*, vol. 37, no. 2, pp. 493-516, 2019.
[9] I. De Miguel *et al.*, "Cognitive dynamic optical networks," *IEEE/OSA J. Opt. Commun. and Netw.*, vol. 5, no. 10, pp. A107-A118, 2013.
[10] J. Li, D. Wang, S. Li, M. Zhang, C. Song, and X. Chen, "Deep learning based adaptive sequential data augmentation technique for the optical network traffic synthesis," *Opt. Exp.*, vol. 27, no. 13, pp. 18831–18847, 2019.
[11] L. Deng, "A tutorial survey of architectures, algorithms, and applications for deep learning," *APSIPA Trans. Signal Inf. Process.*, vol. 3, no. 2, pp. 1–29, Jan. 2014.
[12] C. Zhang, D. Wang, L. Wang, J. Song, S. Liu, J. Li, L. Guan. Z. Liu, and M. Zhang, "Temporal data-driven failure prognostics using BiGRU for optical networks," *IEEE/OSA J. Opt. Commun. and Netw.*, vol. 12, no. 8, pp. 227-287, 2020.
[13] C. Zhang, D. Wang, C. Song, L. Wang, J. Song, L. Guan. Z. Liu, and M. Zhang, "Interpretable learning algorithm based on XGBoost for fault prediction in optical network," in *Optical Fiber Communication Conference* (Optical Society of America, 2020), paper Th1F. 3.
[14] J. Li, D. Wang, M. Zhang, and S. Cui, "Digital twin-enabled self-evolved optical transceiver using deep reinforcement learning," *Opt. Lett.*, vol. 45, no. 16, pp. 4654-4657, 2020.
[15] D. Wang, Y. Song, J. Li, J. Qin, T. Yang, M. Zhang, X. Chen, and A. C. Boucouvalas, "Data-driven optical fiber channel modeling: a deep learning approach," *IEEE J. Lightwave Technol.*, vol. 38, no. 17, pp. 4730-4743, 2020.